\documentstyle[prd,aps,preprint,epsf]{revtex}
\tightenlines
\newcommand {\be}{\begin{eqnarray}}
\newcommand {\ee}{\end{eqnarray}}

\def\bar{\overline}
\def\gsim{\lower0.5ex\hbox{$\stackrel{>}{\sim}$}}
\def\lsim{\lower0.5ex\hbox{$\stackrel{<}{\sim}$}}

\global\firsttabfalse
\global\firstfigfalse

\def\figone
{
\begin{figure}[t]
 \begin{center}
 \leavevmode
 \epsfxsize=0.60\hsize \epsfbox{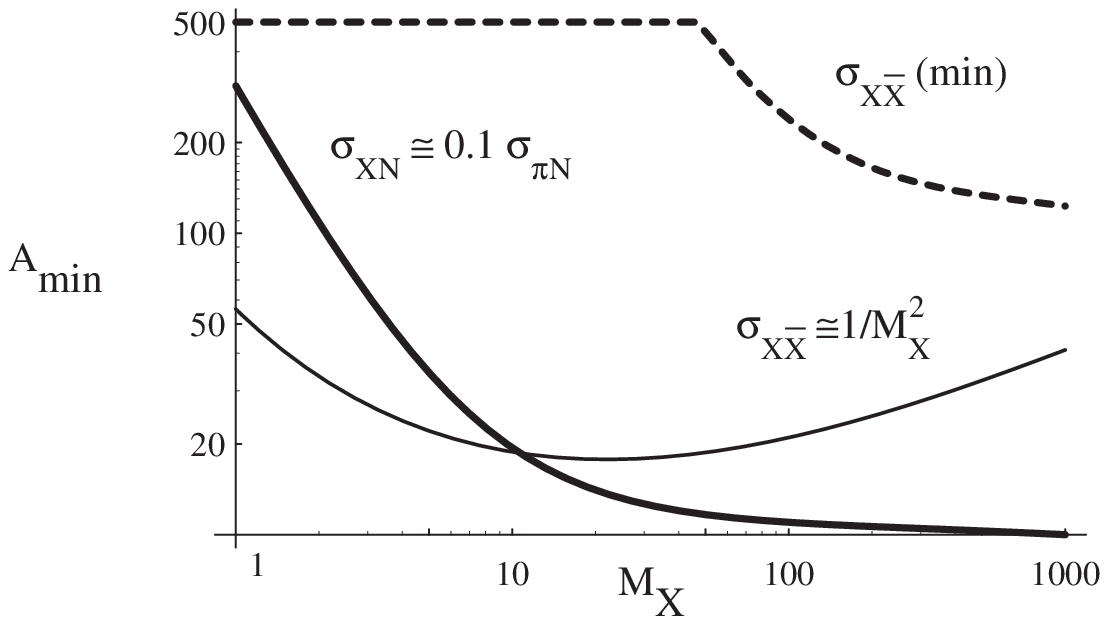}
 \end{center}
 \tightenlines
 \caption{ 
The results of solving equation (\protect\ref{bx2}) 
for three choices for $\sigma_{XN}$. 
The dotted line is for $\sigma_{X\bar{X}} (min)$,
the thin line is for $\sigma_{X\bar{X}}\simeq 1/M_X^2$,
and
the thick line is for $\sigma_{XN} \simeq   \sigma_{\pi N}$.
 }
 \label{fig:one}
\end{figure}
}
\def\figtwo
{
\begin{figure}[t]
 \begin{center}
 \leavevmode
 \epsfxsize=0.60\hsize \epsfbox{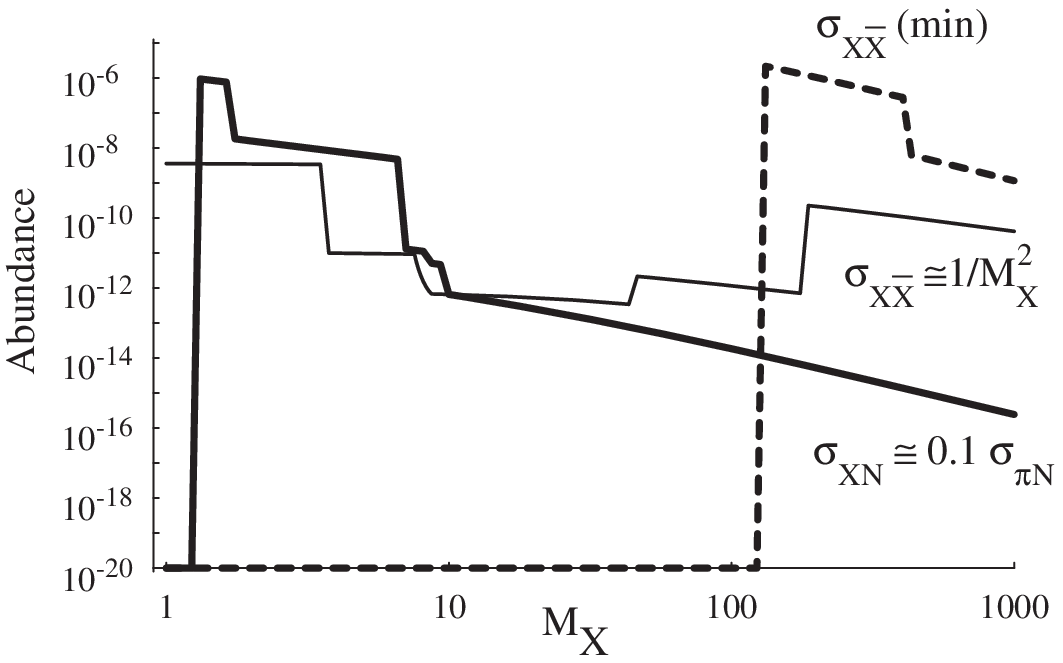}
 \end{center}
 \tightenlines
 \caption{ 
 The  abundance fraction $f$ as a function of $M_X$ for
 the 3 cross section choices.
The dotted line is for $\sigma_{X\bar{X}} (min)$,
the thin line is for $\sigma_{X\bar{X}}\simeq 1/M_X^2$,
and
the thick line is for $\sigma_{XN} \simeq   \sigma_{\pi N}$.
 }
 \label{fig:two}
\end{figure}
}
\def\figthree
{
\begin{figure}[t]
 \begin{center}
 \leavevmode
 \epsfxsize=0.50\hsize \epsfbox{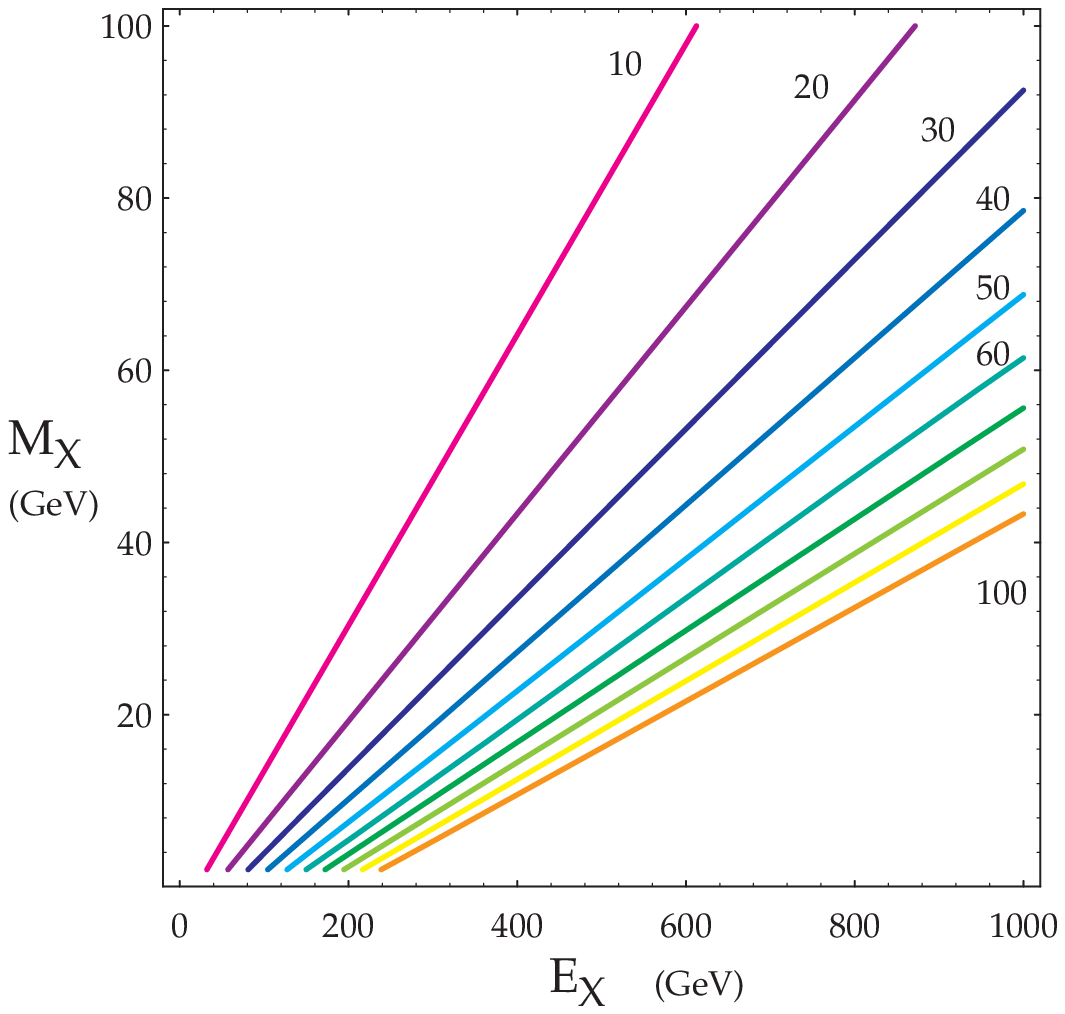}
 \end{center}
 \tightenlines
 \caption{ 
Contours for energy loss as a function of $\{M_X,E_X \}$. 
The contours displayed are in steps of 10 GeV. 
 }
 \label{fig:three}
\end{figure}
}

\begin{document}

\preprint{\vbox{
\hbox{hep-ph/9906421}
\hbox{UMD-PP-99-118}
}}
\title{Searching for Strongly Interacting Massive Particles (SIMPs)}
\author{R. N. Mohapatra$^1$, F. Olness$^2$, R. Stroynowski$^2$, V. L. Teplitz$^2$ }

\address{
$^1$Department of
Physics, University of Maryland, College Park, MD, 20742\\
$^2$ Department of Physics, Southern Methodist University, Dallas, TX 75275}
\date{May, 1999}
\maketitle
\begin{abstract}
We consider laboratory experiments that can detect stable, neutral
strongly interacting massive particles (SIMPs).  
We explore the SIMP annihilation cross section from its minimum value 
(restricted by cosmological bounds) to the barn range, and vary the
mass values from a GeV to a TeV. We calculate,
as a function of the 
SIMP-nucleon cross section,
the minimum nucleon number $A$ for which there
should be binding in a nucleus. 
 We consider accelerator mass spectrometry
with a gold ($A=200$) target, and compute the likely abundance of
anomalous gold nuclei if stable neutral SIMPs exist. We also consider the
prospects and problems of detecting such particles at the Tevatron. We
estimate optimistically  that such detection might be possible for SIMPs
with SIMP-nucleon cross sections larger than 0.1 millibarn and masses between
25 and 50~GeV.
\end{abstract}

\section{Introduction}

The nature of the dark matter in the universe is a question of great
interest for particle physics since the successful standard model has no
candidate with the right properties to qualify as one. It is, therefore,
hoped that determining the properties of the dark matter particle
can provide important information about the nature of new physics beyond
the standard model. One intriguing candidate that has been discussed in 
literature is a stable massive
strongly interacting neutral particle (SIMP)\cite{dimo}. Such particles
arise in many gauge models\cite{nimo}, 
and their existence is  of interest  even if they do not constitute dark matter.
 While there already exist several
severe constraints from cosmology and astrophysics on these
particles,\cite{dimo,PRL} they can still be viable if their strong
interaction parameters are appropriately chosen. It is therefore of
interest to explore all possible avenues to ``tighten the noose'' on
them. In this paper we discuss two experiments of interest
in this connection. 

The essential property of the SIMPs (henceforth denoted as $X$)
that determines their possible experimental
signature is the nature of their interaction with  ordinary nucleons. It is not easy to
guess the interactions of such unknown particles; however, we have some guide
from nucleon-hyperon interactions. 
 In addition,  as detailed in Section~II, we make use of an 
approximate relation between $\sigma_{XN}$ and $\sigma_{X\bar{X}}$.
 Furthermore, cosmology provides a lower bound on $\sigma_{X\bar{X}}$. 
 In considering binding of $X$ in nuclei,   we will assume the XN
potential to be attractive at low energies, as should be the case
whenever $XN$ scattering is a result of two particle exchange.\cite{farrar2}

In a recent publication\cite{PRL} two of us
(RNM and VLT) investigated constraints on SIMPs from searches for anomalous 
nuclei containing them.  We considered two cases.  
 First we built on the
work of Dicus and Teplitz,\cite{dt} assuming that the depth of binding potential $V_{XN}$
is of the order of a few MeV's, as is the case for the $\Lambda$
particles. That work  showed that the anomalous
nucleus ${}^9Be^*_{X}$ would be formed during big bang nucleosynthesis (BBN) if
the  $X$-$N$ potential is attractive and large enough ({\it i.e.},  10\,MeV
or more) for X-binding in light nuclei.  Using a plausible model for the 
$X\bar{X}$ annihilation cross section, Ref.~\cite{PRL} estimated that the
concentration
of the $X$ particles relative to the baryons must be bigger than $\sim 10^{-5}$, 
which then leads to a concentration of ${}^9Be^*_X$ in the present universe 
which is several orders of
magnitude higher than the present upper limits on anomalous $Be$ obtained
in experiments by Hemmick {\it et al.}\cite{hemmick}

Second, Reference~\cite{PRL} pointed out
that, for a
given attractive $X$-$N$ potential, binding
is more likely in the more massive high Z   nuclei than in low Z nuclei.  This
is because, by the uncertainty 
principle, there is less kinetic energy for the SIMP in the high Z
nucleus. 
 We noted, however, the absence of data for
anomalous nuclei with Z
above that of sodium (A=23, Z=11), and urged that
experiments be undertaken in high Z nuclei. 
Since publication of Ref.\cite{PRL}, we have been informed\cite{EPHR} that one group
hopes to look for anomalous gold (A=200, Z=79) nuclei using accelerator mass
spectrometry.

The purpose of this paper is to build upon Ref\cite{PRL} in two directions.
  First we compute $X$ binding and abundance in gold as 
a function of $M_X$ and $\sigma_{XN}$.  We choose gold in view of
the experiment of Ref. [8]. Results for other heavy nuclei will be
similar. Second,
we make an initial estimate for detection of $X$-$\bar{X}$ pairs at 
the Fermilab Tevatron.
The plan of the paper is as follows: in Section~II we estimate the
abundance of 
anomalous gold as a function of $M_X$ and $\sigma_{XN}$ for the region of 
the $\{ M_X,\sigma_{XN} \}$ plane for which there is binding. In Section~III we 
consider Tevatron detection.  Section~IV gives our conclusions.

\section{Searching in Gold}

We perform two calculations in this section.  
 First, we compute   $X$-$A$ binding as a
function of nucleon number $A$ and $M_X$ for a range of $\sigma_{XN}$.  Second, we
compute the abundance of anomalous gold, relative to normal gold, as a
function of $M_X$ and $\sigma_{XN}$.

We consider first the  nuclei in which the X-particle will bind.
As in Ref.\cite{PRL} and works cited therein, we take the  X-A binding
energy to be:
 \begin{equation}\label{bx}
B_{X} = V_{XN}-\frac{\pi^2}{2\mu R^2}
\end{equation}
where 
 $V_{XN}$ is the average X-N potential at rest,
 $\mu$ is the reduced mass of the X-A system, 
 and
  R is the size of the nucleus in which $X$ is bound.
We have used a simple particle in a box model.  
We take $R\simeq R_0\,A^{1/3}$  
with $R_0\simeq 1.4\, fermi$, 
and $A$ is the atomic number of the nucleus.
 The reduced mass of the X-A system is given by the usual expression:
\begin{equation}\label{mu}
\frac{1}{\mu} = \frac{1}{M_A} + \frac{1}{M_X}
\end{equation}

The average X-N  potential, $V_{XN}$,  is defined implicitly via:
\begin{equation}\label{vxn}
\frac{V_{XN} }{V_{NN}} = \sqrt{ \frac{\sigma_{XN} }{ \sigma_{NN} } }
\end{equation}

 If we take the  low  energy NN scattering cross section $\sigma_{NN}$ 
to be on the order of $1\,barn$, 
 use $V_{NN}\simeq 50\,$MeV, 
 and  assume $B_X\ll V_X$, we can write equation (\ref{bx}) as
\begin{equation}\label{}
\frac{2 R_0^2 }{\pi^2}  \ V_{NN}\  \sqrt{ \frac{\sigma_{XN} }{ \sigma_{NN} } }  \   A^{2/3} 
= 
\frac{1}{M_A} + \frac{1}{M_X} 
\end{equation}
which numerically reduces to 
\begin{equation}\label{bx2}
\frac{1}{2 \, GeV} \ 
 \sqrt{ \frac{\sigma_{XN} }{ 1\, barn } } \ A^{2/3}
=
\frac{1}{M_A} + \frac{1}{M_X} 
\end{equation}

\figone

In Figure 1, we give the results of solving equation (\ref{bx2}) 
for three choices for $\sigma_{XN}$. 
 We determine the minimum value 
of A for which there is binding as a function of $M_X$.  
We assume an approximate 
factorization hypothesis of ref.~\cite{PRL}, 
\begin{equation}\label{approx}
\beta \ \sigma_{X\bar{X}} \ \sigma_{NN} \simeq \sigma^2_{XN}
\end{equation}
where $\beta \sim 1$. 

To be definite, we shall consider three values for  $\sigma_{X\bar{X}}$.

\begin{enumerate}

\item
We consider $\sigma_{X\bar{X}} \sim 3\times10^{-13}barn$ as a lower bound. 
This is the  minimum  cross section allowed which does not 
overclose the universe, (assuming  $M_X \gsim$\, GeV).

\item
 We use the estimate 
of ref.\cite{MN},  $\sigma_{X\bar{X}}\simeq 1/M_X^2 = 0.4mb/M_X^2(GeV)$

\item
 Finally we look at the estimate of ref.\cite{farrar},
 $\sigma_{XN}\simeq 0.1\sigma_{\pi N}\sim 2$mb.  
This estimate takes $X$ to be  the particle responsible 
for ultra-high energy cosmic 
ray (UHECR) events\cite{haya}, the UHECRON. 

\end{enumerate}

For case (1), the minimum $\sigma_{X\bar{X}}$,  
we see  in Figure~1 that  gold will only capture X's 
for $M_X\gsim 80$, (assuming the interaction is attractive).
  For case (2), $\sigma_{X\bar{X}} \sim 1/M_X^2$, 
we estimate there is always binding in relatively 
light nuclei, In particular,  much of the $M_X$ range (15$\sim$150 GeV, roughly) 
would have  binding in sodium, and hence is likely ruled out by current 
experiments.
 For case (3), the  UHECRON, there is 
binding in gold for $M_X$ above a few GeV.

We turn now to the less straightforward problem of estimating the abundance 
of anomalous gold.  Our first requirement is a capture scenario.  Two 
possibilities present themselves.  First, the gold on Earth was likely made 
by a supernova in the giant molecular cloud (GMC) in which the sun was bound.
  GMC's live for tens of million of years, so one could assume capture in the 
GMC before formation with an exposure time on the order of $10^7$ years.
There
 are complications, however. The $X$ needs to be slowed from the galactic
virial
 velocity down to thermal
velocities, and then captured.  In addition, 
both the capture reaction, 
 $X+A\rightarrow {}^X A+\gamma$, and the 
dissociation reaction, 
 $\gamma + {}^X A\rightarrow X + A$ from $\gamma$'s 
generated by hot stars and supernova in the GMC, must be considered.  There
appear to be too many unknown parameters for a reliable estimate. The 
second, more conservative and more easily estimated scenario is to assume
SIMP capture after formation of 
the Earth.

The key to the second scenario is that gold nuclei be close enough to the
surface for galactic halo SIMPs to reach them. Geologists say, with some 
confidence,\cite{Herrin} that there are selected gold deposits that
should have had long exposure to any cosmic SIMPs.  These are
deposits found at the surface of the 
earth, in particular, in the gravel of streams (``placer gold") in regions 
that are sufficiently inactive geologically that one can be reasonably
confident
 of an exposure for over $\sim 10^7$ years.  The Sierra Nevada Mountains  
are such 
a region. Placer gold from the California gold rush would have a lesser 
exposure time, on the order of $\sim 10^4$ years. For such placer gold, 
the results discussed below on the abundance of anomalous gold should be 
decreased by a factor of $10^3$.

Given an exposure time of $\sim 10^7$ years, we must then calculate how deep 
into the Earth the $X$ particle penetrates, and how many nuclei are in the 
region penetrated.
  We make the conservative approximation that the gold is uniformly  mixed
with other 
elements so $X$ can be captured by any $A>A_{min}$.  
We assume all X's are captured since
 scattering slows them to thermal velocities.

\begin{table}[t]
\begin{center}
\begin{math}
\begin{array}{||c|c|c|c||} \hline\hline
{\rm Range\  of}\  A & A_M & {\rm Element} & {\rm C_M\ in}\ (\mu gm/ gm) \\ \hline\hline
\leq 20 & 20 & Ar & 5\times 10^5 \\ \hline
10 \leq A \leq 26 & 26 & Al & 8.3\times 10^4 \\ \hline
26 \leq A \leq 137 & 137 & Ba & 425 \\ \hline
137 \leq A \leq 200 & 207 & Pb & 12.5 \\   \hline\hline
\end{array} 
\end{math}
\end{center}
\caption{Abundances of the most abundant nuclei in a given A-interval.} 
\end{table}

We now compute the fraction $f$ of gold that has captured an X:
\begin{equation}
\label{frac}
f(M_X,\sigma_{XN}) \simeq n_X \ v \ t \ \frac{f_{Au}}{N_{Au}}
\end{equation}
Here $n_X$ is the galactic halo abundance at the solar distance from 
the galactic center, 
 $v$ is the galactic virial velocity (300 km),
 $t$ is the exposure time, 
 $f_{Au}$ is the fraction of the stopped X's captured in gold, 
 and
 $N_{Au}$ is the number of Au nuclei (per cm${}^2$) 
 in the stopping length $\lambda_X$ of X.

The fraction  $f_{Au}$  is taken to be: 
\begin{equation}\label{fracx}
f_{Au}  = 
\frac{C_{Au}}{C_M} \ 
\sqrt[3]{ \frac{A_M}{A_{Au}} }
\end{equation}
where the $C$'s are concentrations by weight in the Earth's crust, and we expect
 the capture cross to be proportional to $A^{2/3}$.
The index M designates the most abundant element with a nucleus heavier than
 $A_{min}$. We choose $A_M$ from Table I.

$N_{Au}$ in equation (\ref{frac}), the number of Au nuclei (per cm${}^2$)
in the stopping length $\lambda_X$ of X, is given by
\begin{equation}\label{nau}
N_{Au} = \lambda_X \ 
\frac{M_E}{m_p} \ 
\frac{ C_{Au}}{A_{Au}}
\end{equation}
where $M_E$ is the mass of the earth, 
 and
 $m_p \sim 1$ GeV is the mass of an ordinary nucleon.

For $\lambda_X$ we take
\begin{equation}\label{lam}
\lambda_X = 
\left\{ \frac{\rho_E}{\widetilde{A} m_p} \ 
\widetilde{A}^{2/3} \ \sigma_{XN} \ 
\left[1- \left(\frac{M_X}{M_X + \widetilde{A} m_p}\right)^2\right]
\right\}^{-1}
\end{equation}
where $\widetilde{A} \sim 20$ is the $A$ value for an ``average nucleus,"
 and  the  last factor $(\equiv \Delta E_X/E_X)$ assumes that $XA$ scattering 
in the C.M. is isotropic.  
Note, as a sanity check, that the $C_{Au}$ factor in equation(\ref{nau}) 
cancels with that in $f_{Au}$, equation(\ref{fracx}), as one would expect.

Finally, to get $n_X$ in equation(\ref{fracx}) we compute the primordial 
X freeze-out abundance, $\tilde{n}_X$, precisely as specified in 
Kolb and Turner\cite{K and T}, using  (in 
their notation) n=1 and g=1, and making use of the factorization
approximation relating $\sigma_{X\bar{X}}$ to $\sigma_{NX}$. 
We then assume that X's 
are concentrated in the galaxy to the same extent as baryons, but are 
distributed in an isothermal spherical halo of 70\,kpc radius.  This gives 
$n_X/n_B=3.7\times10^{-3}\tilde{n}_X/\tilde{n}_B$ with, the tilde denoting
the cosmic average.

The result of the calculation for  log${}_{10}(1/f)$ 
is shown with quiet drama in Table 2.  
$\sigma_{XN}$ ranges from $5\times 10^{-7} b$ 
(corresponding to $\sigma_{X\bar{X}} = \sigma_{min}=3\times 10^{-37}cm^2$) 
to 1 barn,   while $M_X$ ranges from 1 GeV to 1 TeV.  
 In the first column we see that the $X$ does not bind in gold 
until  $M_X\in[80,130]$ GeV, which is consistent with Figure~1.  
As $\sigma_{XN}$ increases, 
 the abundance falls because\cite{K and T} 
$Y_{\infty}\propto 1/(M_X \sigma_{X\bar{X}})$ and
 $\sigma_{X\bar{X}}\sim \sigma_{NX}^2$. This fact also plays a major role
 in the decrease of $f$ as $M_X$ increases.
Note from the table that the boundary between binding and
no-binding can be approximated by the implicit relation
$M_X^2\sigma_{XN} \sim 5mb\, {\rm GeV}^2$.

The result of Table 2 is optimistic  in that previous
 experiments have reached abundance fractions approaching $f \sim 10^{-20}$. 
Therefore, it is likely to be possible to explore the entirety of the 
 parameter space of Table~2 in which there is binding of the X. For the
 UHECR case of Ref.\cite{farrar} $(\sigma_{XN}\gsim 0.1\sigma_{\pi N},
 M_X\lsim 50\,$GeV), it is only necessary to go to $10^{-16}$, not
$10^{-20}$.  This is because Table 2 shows that, even for the largest XN
cross sections, if the value of $M_X$ is below 50 GeV, log${}_{10}(f)>-15.7$.

While Table~2 shows the fraction $f$ is a relatively smoothly varying function of 
 $\sigma_{NX}$ and $M_X$ over most of the range, 
 closer examination shows some regions  of large
 variation.  
For example, the first column shows a change by 2 orders of magnitude 
between  $M_X$=350 and 570 GeV.  
 In Figure~1 this corresponds to $M_X\simeq 530$ where  
$A_{min}$ goes from above 137 to below 137.  This corresponds to $A_M$ in
 equation(\ref{bx2}) falling from 207 (Pb) to 137 (Ba) which shifts 
 $C_M$ in equation(\ref{fracx}) by a factor of 35.
 To see this effect more
 clearly, we plot in Figure~2  abundance fraction $f$ as a function of $M_X$ for
 the 3 cross section choices of Figure~1.  The curve for  $\sigma_{X\bar{X}} (min)$ has a
 steep rise just past $M_X=100$ where $A_M$ in Figure~2 first falls below
 200 (so that there is binding in gold).  The discontinuous behavior at
 $M_X=530$ GeV is seen clearly.  
The curve for  $\sigma_{XN} \simeq   \sigma_{\pi N}$
 is monotonic (after its initial 
rise when binding occurs) corresponding to the monotonic behavior of Au in
 Figure~1 for the corresponding curve.   
 The curve for
$\sigma_{X\bar{X}}\simeq 1/M_X^2$  has
 discontinuities of both signs corresponding to the fall of this same curve in 
Figure~1 to below 26 and 20, and subsequent rise above these points.

\begin{table}[t]
\begin{center}
\begin{math}
\begin{array}{||c||c|c|c|c|c|c|c|c|c|c|c|c|c|c|c||}  \hline\hline
& 0.0005  & 0.0015  & 0.0042   & 0.012  &  0.032 & 0.09 & 0.25 & 0.69 & 
1.9 & 5.3 & 15 & 41 & 110 & 310 & 860  \\  \hline\hline
1.0 & - &  - &  - &  - &  - &  - &  - &  - &  - & 6.3 & 8.3 & 8.7 & 12.5 & 12.9& 13.4 \\ \hline
1.6 & - &  - &  - &  - &  - &  - &  - &  - & 6.1 & 8.1 & 8.5 & 12.3& 12.7 & 13.1 & 13.6 \\ \hline
2.7 & - &  - &  - &  - &  - &  - &  - & 5.9 & 7.9 & 8.3 & 12.1 & 12.5 & 12.9 & 13.3 & 13.8\\ \hline
4.3 & - &  - &  - &  - &  - &  - & 5.7 & 7.7 & 8.1 & 11.1 & 12.3 & 12.7 & 13.1 & 13.6 & 14.0\\ \hline
7.1 & - &  - &  - &  - &  - & 5.5 & 7.5 & 7.9 & 10.9 & 12.1 & 12.5 & 12.9 & 13.4 & 13.8 & 14.2\\ \hline
12  & - &  - &  - &  - & 5.6 & 7.6 & 8.1 &8.5 & 12.2 & 12.7& 13.1 & 13.5 & 13.9 & 14.3 & 14.8\\ \hline
19  & - &  - &  - &  - &7.5 & 7.9 & 8.3 &11.3 & 12.5 & 12.9 &13.3 & 13.8 & 14.2 &14.6 & 15.0\\ \hline
31  & - &  - &  - & 7.4 &7.8 &8.2 & 8.6 & 12.4&12.8 & 13.2 & 13.6& 14.1 & 14.5 & 14.9 & 15.3\\ \hline
50  & - &  - & 5.7 &7.7 &8.1&8.5 &11.5& 12.7 & 13.1 & 13.6 & 14.0 &14.4 &14.8 &15.3 &15.7\\ \hline
81  & - &5.7 &7.7 &8.1 &8.5 &8.9 &11.9 &13.1 &13.5 &14.0 &14.4 &14.8 &15.2 &15.6 &16.1\\ \hline
132 & 5.7&7.7 &8.1 &8.5 &8.9 &9.3 &12.2 &13.5 &13.9 &14.3 &14.7 &15.2 &15.6 &16.0 &16.4\\ \hline
220 & 6.0 &8.0 &8.4 &8.9 &9.3 &9.7 &12.6 &13.9 &14.3 &14.7 &15.1 &15.5 &16.0 &16.4 &16.8\\ \hline
350 & 6.4 &8.4 &8.8 &9.3 &9.7 &10.1 &13.8 &14.3 &14.7 &15.1 &15.5 &15.9 &16.4 &16.8 &17.2\\ \hline
570 & 8.4 &8.8 &9.2 &9.7 &10.1 &10.5 &14.3 &14.7 &15.1 &15.5 &15.9 &16.4 &16.8  &17.2 &17.6\\ \hline
930 & 8.9 &9.3 &9.7 &10.1 &10.5 &10.9 &14.7 &15.1 &15.5 &16.0 &16.4 &16.8 &17.2 &17.6 &18.1\\  \hline\hline
\end{array}
\end{math}
\vskip 10pt
\caption{
$M_X$ (vertical) is in units of GeV, and $\sigma_{XN}$ (horizontal) is in units of $mb$.  
 Table entries are $\log_{10}(1/f)$, and
 the $-$ indicates those cases for which $X$ does not bind at all.
}
\end{center}
\end{table}

\figtwo

It is also of interest to make a connection with the experiments searching for
weakly interacting massive particles (WIMPs).  
For $\sigma_{XN}\gsim 4\times10^{-6}b$, which corresponds to 
$\sigma_{X\bar{X}}$ a factor of 50 above $\sigma_{X\bar{X}}$(min), the 
contribution of $X$ to the local galactic dark matter 
density ($4\times10^{-25}g/cm^{3}$) is less than 2 percent.

It is important to note that  we only assume that the
mass density of SIMPs  is less than or equal to the
dark matter constraint; we do not assume, as did 
Starkman {\it et al.}, \cite{dimo},  that it saturates the
constraint. 

In summary, we see from the above discussion that: 
 i) we have binding in gold
 for $M_X^2\sigma_{XN} \gsim 5mb\, {\rm GeV}^2$; 
 ii) a SIMP of mass
 up to a TeV satisfying this condition, and having an attractive interaction
 with nucleons, should form anomalous gold nuclei of sufficient abundance
 to be seen by accelerator mass spectrometer experiments sensitive to one
 part in $10^{20}$; and 
 iii)  UHECRs could be detected in an experiment with a sensitivity of
 one part in $10^{16}$ for the mass range from a few GeV to their 50 GeV
maximum, (assuming a target of gold with $10^7$ years' exposure).

\section{Searching at the Tevatron}

If neutral, stable SIMPs were actually to exist, it might be possible to
produce and  detect them at the Tevatron.
Although such detection might be difficult,   further study is required
before one can decide whether it is impossible.
In this section, we  make an optimistic
estimate of the most promising ranges of $M_X$ and $\sigma_{XN}$ for observing
these SIMPs.  
 While the region accessible to the Tevatron is quite restricted, 
it includes a portion of  a particularly interesting region -- the 
region relevant to the explanation of
 the Ultra High Energy Cosmic Ray (UHECRs) events proposed by 
 Farrar, Kolb and coworkers.\cite{farrar}

We consider $X \overline{X}$ production in one of the Tevatron detectors.
Assuming that the SIMP has colored constituents 
({\it e.g.},  gluon, gluino, ...), its
pair production will be accompanied by soft hadrons.  
 A neutral, stable SIMP particle will not generate a signal in the central tracker,
and is unlikely to interact in the electromagnetic calorimeter of
the Tevatron detectors. It may, however, interact with  material
in the denser hadronic calorimeter. Such an interaction will have  kinematical
characteristics that can be used to distinguish it on a statistical basis
from neutron and $K^0_L$ interactions.
The distinction is based on the
assumption that at the Tevatron energies, the mass of a SIMP represents
a substantial fraction of its total energy.
 We then need to ask how
much energy the SIMP is likely to deposit in the hadron calorimeter,
whether we can distinguish a SIMP shower from that from a neutron or
$K_{L}^0$, and what overall event rate might be expected.

\figthree

With regard to energy deposition in the hadron calorimeter, we make a
qualitative estimate.  Assume that, in interacting with a nucleon in one
of the steel plates, the $X$ loses half of its C.M. kinetic energy into
particle production.  We can then transform back to the lab system and
estimate the energy lost by $X$ as a function of $E_X$ and $M_X$.  Figure~3
shows the results.  It has contours for energy losses (into hadron
showers) of 10, 20,... 100 GeV. 
  We see that  400 GeV partons could,  triggering on multiple $\sim$10 GeV showers,
permit discovery of SIMPs with masses up to 60 GeV.  Similarly 600 GeV
partons could push the SIMP discovery limit up to 100 GeV. At 300 GeV,
we can get up to 50 GeV masses, which is the upper limit, in the
analysis of Farrar et al \cite{farrar}, of a SIMP that would explain
the UHECR events.
 
If $\sigma_{XN}$ is too small, there will be no showers at all.  
In $1m$ 
 of steel, we get about 5 interactions for a cross section of a few $mb$, but
only $0.001$ for our minimum $\sigma_{XN}$ ($\sim 5\times 10^{-31} cm^2$).
We might detect  one jet from the SIMP pair  
at a sufficient rate if we had a cross section on
the order of at least a tenth of a millibarn, 
and both jets for a cross section over a millibarn.
 The estimate of reference~\cite{farrar}
states that a cross section of more than a millibarn
is needed if the SIMP is to reproduce the UHECR events.

We estimate the production rate of SIMPs by scaling  the production
rate of jets by the ratio of $XN$ to the Meson-N cross sections.
We observe that the total cross section for $\pi p$ and $K p$ scattering
is approximately $\sim 30 mb$ for $\sqrt{s}$ of a few hundred GeV, with
a mild (logarithmic) variation with $\sqrt{s}$.
 With this perspective, we can now turn from the $t$-channel $Xp \to Xp$ 
process to the $s$-channel $p \bar{p} \to X \overline{X}$ to estimate a 
scaled cross section.  In the low energy region, X-production will be
suppressed
relative to  jet production by the $X$ mass. However, for 
$E_X$ larger than a few times $M_X$, there is no more phase space 
suppression, and we therefore expect scaling to work well 
(for $M_X=50\, $GeV) in the
region $E_X \gsim 200 GeV$. The cross section for producing jets with
$E_T\gsim 200 GeV$ is  $\sim 100 pb$.  This gives  $\sim 3 pb$ for any one
quark (to be compared with a t-quark production cross section of $\sim 5 pb$), 
corresponding to  $\sim$300 events in Run~I and $\sim$6000 events in 
Run~II.  Again, reference~\cite{farrar} estimates that the UHECR would need a
cross section of about a tenth the meson nucleon cross section.  Scaling
the production cross section by that amount gives a good number of
events in Run~II.

Turning to the question of whether we would recognize a SIMP shower in a
hadron calorimeter if we saw one, the best available discriminant would
appear to be the opening angle of the shower.  A pion moving
transverse in the SIMP-nucleon C.M. system will have a laboratory angle
of 
\begin{equation}
\tan \theta = \frac{1}{\gamma} =
\frac{ \sqrt{2  m_p E_X +M^2_X} }{E_X+m_p }
\end{equation}
 Comparing a  500
GeV SIMP with a  500 GeV neutron, we see that the SIMP shower will be
$\sim$ 40 percent wider if $M_X \gsim 25 GeV$.

In summary, an optimistic scenario finds the Tevatron discovery
potential in the range $\sigma_{XN} \gsim  0.1mb$ and 
$25 GeV \lsim  M_X \lsim  50
GeV$.  
Consequently, it might be possible to resolve the UHECR puzzle at Fermilab.

\section{Conclusions}

We emphasize here that, as discussed in Ref.~\cite{PRL}, searches for
SIMPs in anomalous nuclei are much better carried out in high Z nuclei.
In such nuclei, the SIMP kinetic energy is minimized so that the chances
of binding are maximized.  Table~2 shows that a proposed
accelerator mass spectrometry experiment in gold is capable of discovering
or eliminating SIMPs over $\sim$80 percent of the relevant portion of the
$(\sigma_{XN},M_X)$ plane so long as the low energy X-N potential is
attractive, (the likely event).\cite{farrar}

Looking for $X$ at the Tevatron is difficult.  One needs events in which
there are no high energy particles at the vertex or in the electromagnetic
calorimeter, but there are showers in the hadron calorimeter.  The range of
$M_X$ values for which the hadron calorimeter showers are sufficiently
energetic goes up to  50 or  100 GeV for the $M_X$ mass.  The range for
which the shower opening angles are sufficiently wide to distinguish them
from neutron or Kaon showers  begins around 25 GeV for the $M_X$
mass.  Although this region is limited, it includes at least half the
region of interest for a possible UHECR explanation.\cite{farrar}

Acknowledgments: We thank 
D. Berley, 
K. Brockett,
K. De,
D. Dicus, 
M.A. Doncheski, 
R. Ellsworth, 
G. Farrar, 
E. T. Herrin,
D. Rosenbaum,
R. Scalise,
and
G. Yodh.
 The work of RNM has been supported by
the National Science Foundation grant under no. PHY-9802551.  
The work of Olness, Stroynowski, and Teplitz is supported by DOE.



\begin{thebibliography}{99}
\bibitem{farrar2}  
We appreciate a conversation with Professor Farrar in
which she emphasized the fact that two particle exchange dominance makes
XN scattering attractive rather generally.
\end{thebibliography}
\end{document}